# Designing a Real-Time IoT Data Streaming Testbed for Horizontally Scalable Analytical Platforms: Czech Post Case Study

Martin Štufi[1][a] and Boris Bačić[2][b]

[1]Solutia, s.r.o., Prague, Czech Republic
[2]School of Engineering, Computer and Mathematical Sciences, Auckland University of Technology, Auckland, New Zealand
martin.stufi@solutia.cz, boris.bacic@aut.ac.nz

Abstract: There is a growing trend for enterprise-level Internet of Things (IoT) applications requiring real-time horizontally scalable data processing platforms. Real-time processing platforms receiving data streams from sensor networks (e.g., autonomous and connected vehicles, smart security for businesses and homes, smartwatches, fitness trackers, and other wearables) require distributed MQTT brokers. This case study presents an IoT data streaming testbed platform prepared for the Czech Post. The presented platform has met the throughput requirement of 2 million messages per 24 hours (comprising SMS and emails). The tested MQTT broker runs on a single virtual node of a horizontally scalable testbed platform. Soon the Czech Post will modernise its eServices to increase package deliveries aligned with eCommerce and eGovernment demands. The presented testbed platform fulfils all requirements, and it is also capable of processing thousands of messages per second. The presented platform and concepts are transferable to healthcare systems, transport operations, the automotive industry, and other domains such as smart cities.

## 1 INTRODUCTION

The global Internet of Things (IoT) market size had an estimated volume of USD 308,97 billion in 2020. According to Insights (2021), the annual growth projection is 25,4% for the following market period of 2021-2028.

Due to the rapid growth of IoT and sensor networks combined with increasing demands for data exchange and processing, big data analytical platforms play a significant role in enabling infrastructures (Anshu & Yogesh, 2016; Nasiri, Nasehi, & Goudarzi, 2019; Strohbach, Ziekow, Gazis, & Akiva, 2015). As common knowledge today, IoT networks exchange telemetry and other sensor-specific data with devices on-premises or via cloud edge gateways. One of the essential protocols for this type of message-based communication is MQTT (Message Queuing Telemetry Transport, ISO/IEC 20922). Furthermore, for modern societies requiring affordable business solutions involving increasingly large data exchanges and processing demands, horizontally scalable big-data analytical platforms can help overcome technology adoption barriers (Cortés, Bonnaire, Marin, & Sens, 2015; Mavromatis & Simeonidou, 2020). Moreover, big data analytical architectures and IoT sensors have been embraced in almost all fields of society (Ed-daoudy & Maalmi, 2019). As one of its kind in the EU, a horizontally scalable big data analytical platform has been implemented as a nationwide modern healthcare eSystem in the Czech Republic (Štufi, Bačić, & Stoimenov, 2020). However, in the authors' view, the next generation of healthcare systems will also be able to process data streams from sensor networks, including medical IoT devices and various wearable sensors, while at the same time preserving the end-users' privacy.

Making technology selections and designing an IoT testbed for analytical platforms can be challenging tasks. For example, there is a growing number of clustered and non-clustered MQTT brokers available. However, there is also a lack of benchmarks for such messaging brokers, especially for IoT infrastructures. In the absence of publicly available benchmarks, one of the outcomes of this research is to establish performance evaluation criteria. The other outcome of this research focuses on establishing an infrastructure that can guarantee the processing of a number of messages in a given period or messaging throughput performance during data exchange peaks). Furthermore, to identify possible challenges of streaming data processing on the proposed platform.

[a] https://orcid.org/0000-0002-0763-8360
[b] https://orcid.org/0000-0003-0305-4322

## 1.1 Research Questions

To produce an enterprise-grade modern analytical platform for the Czech Post (the industry client), we had to consider the prerequisite messaging throughput requirements and identify possible challenges associated with the topic of investigation. Hence, we should answer the following research questions:
1. Can we design a horizontally scalable system that can process 2 million messages per 24 hours from IoT sensor network devices in a second? Can such a system handle the peaks in order of few thousand IoT messages per second on a single node MQTT broker?
2. Can we propose a minimum viable platform for real- or near real-time streams data processing?
3. How to use a multi-thread Java application for IoT data messaging simulation scenarios and prepare a TestBed Platform (TBP) benchmark?

As a critical component for the intended testbed architecture and analytical platforms development, we present the IoT network simulation concept with developed source code in Java. The multithreading Java application (Figure 1) can send messages in order of thousands per second for a real-time data streaming scenario for a viable architecture.

This platform would also provide real-time data processing with low latencies, high throughput, and a secure manner among horizontally scalable platform components.

This paper is structured as follows: first, it presents the background of IoT streaming data. Section two (Materials and Methods) provides the exact specification of the experimental setup and testbed design. Section three reports benchmark results. Section four discusses limitations in the research scope alongside additional findings and insights. Section five summarises the achieved outcomes and opportunities for future work. Section four Information about the platform implementation.

## 1.2 Background

Regarding IoT standard messaging protocol, one of the early surveys of enabling technologies for industrial production and decision support systems involved stress-testing of MQTT brokers that are capable of connecting large numbers of devices (Al-Fuqaha, Guizani, Mohammadi, Aledhari, & Ayyash, 2015).

Four years later, Bertrand-Martínez, Feio, Nascimento, Pinheiro, and Abelém (2019) have qualitatively evaluated different MQTT brokers according to ISO 25010 quality criteria, among them EMQX and Mosquito. On the other hand, Koziolek, Grüner, and Rückert (2020) compared three representatives, distributed MQTT brokers by using their own evaluation criteria.

Akanbi and Masinde (2020) presented the application of a distributed stream processing framework for the real-time big data analysis of heterogeneous environmental management and monitoring data using Apache Kafka in the Confluent Platform. They demonstrate the suitability and applicability of applying big data techniques for real-time processing and analysis of environmental data

```
private void fillBasicMessages() {
    basicMessages.add(
        new Message( message: "{\"battery\":\"99.5\",\"battery_low\":\"false\",\"contact\":\"false\",\"linkquality\":\"22.62\",\"tamper\":\"false\",\"voltage\":\"952.32\"}",
            topic: "slt/moskevska/arrow/door/door_01_m/rlsb8124b002342cddc/state", retained: false));
    basicMessages.add(new Message( message: "{\"battery\":\"51.12\",\"battery_low\":\"true\",\"contact\":\"true\",\"linkquality\":\"32.06\",\"tamper\":\"false\",\"voltage\":\"3011.03\"}",
            topic: "slt/moskevska/corridor/door/door_main_m/rlsFCCCCFFFF/state", retained: false));
    basicMessages.add(new Message( message: "{\"battery\":\"97.11\",\"humidity\":\"31.51\",\"linkquality\":\"37.35\",\"temperature\":\"10\",\"voltage\":\"2479.62\"}",
            topic: "slt/moskevska/reception/sensor_temp_json/temp_hum_01_m/rlsb0124b00239de24c/state", retained: false));
    basicMessages.add(new Message( message: "{\"battery\":\"13.45\",\"humidity\":\"52.42\",\"linkquality\":\"22.12\",\"temperature\":\"48.3\",\"voltage\":\"2721.81\"}",
            topic: "slt/moskevska/scoop/sensor_temp_json/temp_hum_02_m/rlsb0124b0823a5278f/state", retained: false));
    basicMessages.add(new Message( message: "{\"co2\":\"5096.81\",\"heat_index\":\"9.44\",\"humidity\":\"51.41\",\"pressure\":\"1801.71\",\"temperature\":\"-2.13\",\"tvoc\":\"6791.39\"}",
            topic: "slt/moskevska/boss/sensor_co2/air_quality_sensor_02_m/rlsBTT44ea06ca703T/state", retained: false));
    basicMessages.add(new Message( message: "{\"co2\":\"8072.28\",\"heat_index\":\"30.28\",\"humidity\":\"55.8\",\"pressure\":\"784.57\",\"temperature\":\"30.28\",\"tvoc\":\"5578.04\"}",
            topic: "slt/vrsovicka/crunch/sensor_co2/air_quality_sensor_01_m/rlsFDDDDFFFF/state", retained: false));
    basicMessages.add(new Message( message: "FFFFAAAAFFFF;;2021-08-03 17:03:59;-11.11;;56.67;-7.86;;28.9;;",
            topic: "slt/vrsovicka/outside/sensor/temp_2_m/rlsb40e78e0f309/state/FFFFAAAAFFFF0/state", retained: false));
    basicMessages.add(new Message( message: "dev7e2055baf309;;2021-08-03 17:04:39;30.17;;56.06;;27.08;;39;;",
            topic: "slt/vrsovicka/drill/sensor/temp_1_m/rls7e2055baf309/state", retained: false));
    basicMessages.add(new Message( message: "MOVE_TO_TARGET#29:3",
            topic: "slt/vrsovicka/mainfloor1/agv/develop3/65DB91DEDC52/order", retained: false));
    basicMessages.add(new Message( message: "65DB91DEDC52;;ON_WAY;;Moving;;[20, 13]2021-08-03 at 17:04:28",
            topic: "slt/vrsovicka/mainfloor2/agv/develop3/65DB91DEDC52/state", retained: false));
}
```

Figure 1: Java application "mqtt-performance: A code snippet generating IoT sample test message data (https://github.com/stufim/mqtt-performance).

from heterogeneous systems, contrary to widely adopted Extraction, Transformation, and Loading (ETL) techniques.

Al-Fuqaha et al. (2015) presented an overview of the foundations of this concept, which is enabling technologies, protocols, applications, and the recent research addressing different aspects of the IoT. This, in turn, provides a good foundation for researchers and practitioners interested in gaining insight into the IoT technologies and protocols to understand the overall architecture and role of the different components and protocols that constitute the IoT.

From the bibliographic surveys since 2015, there are not many studies to address the problems of collecting, processing, and analysing large amounts of data in real-time, nor they provide established benchmarking testbed platforms relying on the use of IoT devices or other networked sensors. To address the gap in literature regarding lack of the testbed platforms, this research is focused to implement novel benchmark testbed concept with included underlying architecture, streaming framework and experimental evaluation based on a large number of generated IoT messages per second.

## 2 MATERIAL AND METHODS

The dataset used for this experimental study includes the data generated and submitted by multi-thread Java application "mqtt-performance" to HiveMQ message broker shared as an open-source software on GitHub with its SHA256SUM code signature (Figure 1, Figure 3 and Figure 5). As part of the application design, the IoT test message combines nine messages in one data set. This data set sends an "mqtt-performance" application established on the predefined number of messages, threads with predefined delays between every single data set in the Java class (Figure 3). Also, the application has predefined delays between transmitting every single data set (Figure 3, Java class PublishThread.java). The main java application sends these messages to the predefined IP address of the HiveMQ message broker on default port 1883.

Figure 1 shows a code snippet of a Java application that can benchmark HiveMQ broker in different domains with IoT devices. These devices are, for example, IoT sensors, Zigbee or any sensors and actuators publishing telemetry data to the edge gateway cluster.

Java application implements the MQTT protocol (OASIS, 2019) as the IoT standard messaging and data exchange protocol. Project Object Model file (pom.xml) defines the required dependencies with hivemq-mqtt-client, version 1.2.2 associated to HiveMQ MQTT broker.

Hence, for the high-performance TBP (Figure 2), we have multiple nodes on a layer after HiveMQ ("HiveMQ," 2021) for Apache Kafka ("Apache Kafka," 2021), Apache Spark ("Apache Spark," 2021), Apache Hadoop ("Apache Hadoop," 2021), NoSQL database ("Vertica," 2021).

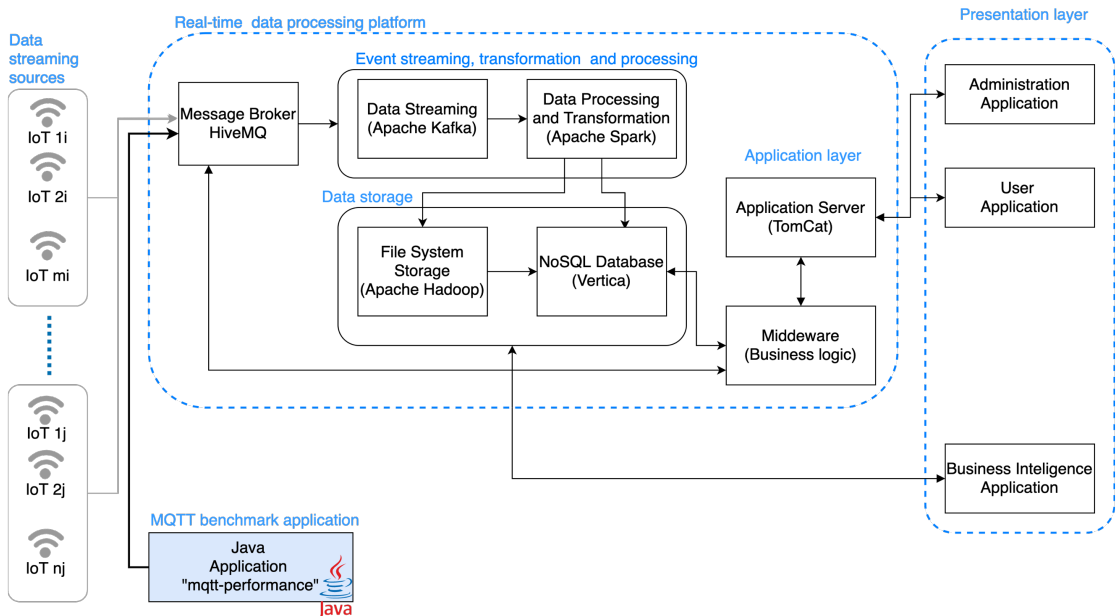

Figure 2: Testbed design: Java Application "mqtt-performance" and real-time processing platform architecture.

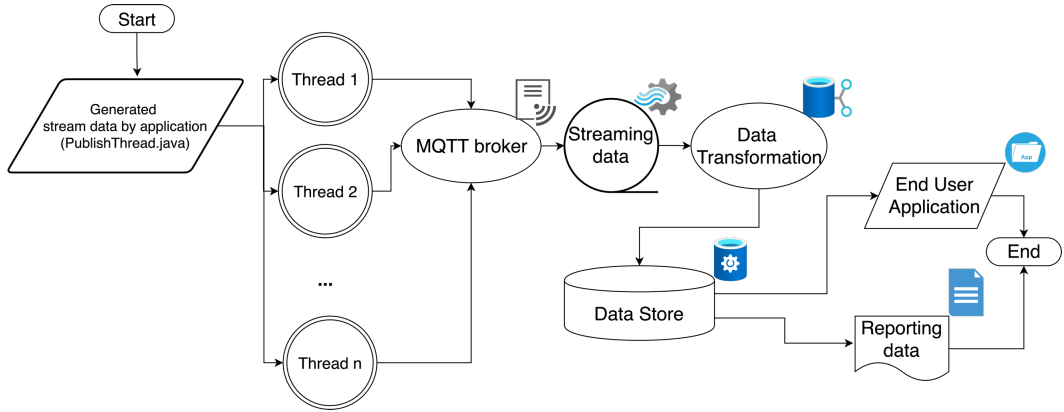

Figure 3: Data-flow diagram https://github.com/stufim/mqtt-performance/blob/master/src/main/java/performance/PublishThread.java.

Client-based messaging application for high-speed, reliable, and efficient handover ensures the data transmitted to the HiveMQ messaging broker using the MQTT protocol. In addition, HiveMQ broker allows automatic elastic and linear scalability at runtime and as such, it is considered a fault-tolerant and resilient cluster, with the properties known as zero-downtime upgrade, and up to ten million supported MQTT clients per cluster (Koziolek et al., 2020).

Figure 2 shows the overall architecture design for the testbed horizontally scalable analytical platform. Instead of data-streaming sources based on IoT sensors or ZigBee, we use the Java-based multithreading application (Solutia, 2021). This application executes many threads that can run concurrently.

## 3 RESULTS

The results of a system prototype confirm that the proposed system can meet the requirements of 2 million messages per 24 hours and achieve approximately 5.000 messages throughput per second during peak times. In addition, the presented system can also process IoT-generated messages from various devices. Thus, our system is also applicable to data processing from smartwatches, wearables, IoT-based devices for near-future healthcare, and the Internet of Robotic for automotive and other service industries.

The presented solution has been implemented as a generic development platform derived from the big-data analytical platform in healthcare (Štufi et al., 2020). The testbed platform shows how effectively to proceed with a few thousand IoT messages in a second. The test datasets were created manually based on a real-data sample set (Solutia, 2021) generated by IoT devices. We used BMP280 Digital Pressure Sensor ("BMP280 Digital pressure sensor," 2021), CCS811 Ultra-low power digital gas sensor for monitoring indoor air ("CCS811 Datasheet," 2021), Si7021-A20 I2C humidity and temperature sensor ("Si7021-A20," 2021). Since the Si7021-A20 sensor is not accurate enough, we used the DHT 11 sensor ("DHT11 humidity & temperature sensor," 2021) together with the air quality sensor. The air quality sensor and the DHT 11 sensor are on one microcontroller and work simultaneously.

Intentionally, during the benchmark of the proposed testbed platform, we performed 16 cold restarts. Based on that, we also make sure to clear the caches including RAM contents as well as performed the boot sequence from scratch.

For automating computer application deployment an open-source container-orchestration platform, scaling, or managing workload in the designed and proposed platform we use Kubernetes (K8s). Hardware specification for testbed platform with the number of nodes, memory, or virtual CPU (vCPU) are shown in Table 1, Table 2 and Table 3.

Table 1: Hardware specification for tested real-time data processing platform.

|  | MQTT broker | Kafka connector | Kafka Zookeeper | Kafka |
|---|---|---|---|---|
| Number of nodes | 1 | 3 | 3 | 3 |
| Memory (GB) | 4 | 4 | 4 | 4 |
| vCPU | 3 | 1 | 1 | 1 |

Table 2: Hardware specification for tested real-time data processing platform.

|  | Spark driver[*] | Spark executor[**] | Hadoop[***] | Vertica |
|---|---|---|---|---|
| Number of nodes | 2 | 4 | 1 | 1 |

|  | Spark driver* | Spark executor** | Hadoop*** | Vertica |
|---|---|---|---|---|
| Memory (GB) | 1.3 | 8 | 16 | 1024 |
| vCPU | 3 | 3 | 4 | 1024 |

Note:
*Spark has two types of nodes; the driver is just for redistributing messages, while the executor processes them. Spark driver has two nodes that are not clustered and are separate applications. Each driver listens and handles different topics.*
**Each spark driver has two executors.*
***Hadoop is not part of the Kubernetes cluster (four nodes in one master). Instead, it runs on a separate VMware server.*

Table 3 shows the Kubernetes (K8s) cluster specification on which we run all components related to TBP.

Table 3: Hardware specification for Kubernetes cluster.

|  | Master node | Workers |
|---|---|---|
| Memory (GB) | 8 | 48 |
| vCPU | 4 | 12 |

Table 4: Testbed performance results on the proposed real-time analytical platform for 50.000 messages.

| Test parameter | Results |
|---|---|
| Messages per second | 5.270,5 |
| The overall number of messages | 50.000 |
| Number of different devices | 10 |
| Message size | 19 – 116 chars |
| Overall duration [ms] | 9.486,8 |
| Batch | 10 |
| Sleep [ms] | 950 |

Table 5: Testbed performance results on the proposed real-time analytical platform for 100.000 messages.

| Test parameter | Results |
|---|---|
| Messages per second | 5.024,3 |
| The overall number of messages | 100 000 |
| Number of different devices | 10 |
| Message size | 19 – 116 chars |
| Overall duration [ms] | 19.903,2 |
| Batch | 20 |
| Sleep [ms] | 950 |

Table 6: Testbed performance results on the proposed real-time analytical platform for 150.000 messages.

| Test parameter | Results |
|---|---|
| Messages per second | 4.937,9 |
| The overall number of messages | 150.000 |
| Number of different devices | 10 |
| Message size | 19 – 116 chars |
| Overall duration [ms] | 30.377,4 |
| Batch | 30 |
| Sleep [ms] | 950 |

Benchmark shows that we generated approximately 5.000 messages per second (Table 4, Table 5, and Table 6). These messages we addressed to the one virtual node of the MQTT broker, with 50, 100 or 150 thousand overall messages sent through the TBP with near-linear throughput performance (Figure 4). With the fixed number of different devices and message sizes, we have the different overall duration for messaging processing TBP from 9,5 seconds to 30 seconds.

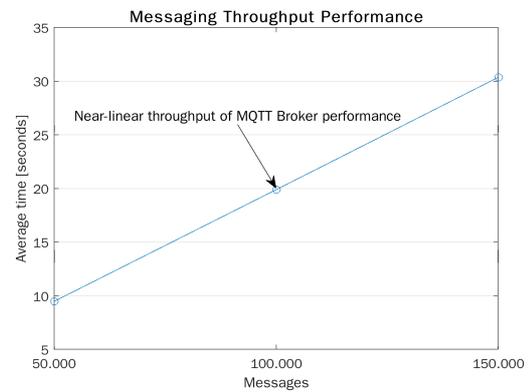

Figure 4: Messaging throughput performance with near-linear through performance on MQTT broker.

During the benchmark testing, we found MQTT broker's decreasing performance. Instead of starting to slow down its processing, MQTT broker starts to slow down IoT message deliveries. These findings suggest that the final target platform should allow horizontal scaling i.e. incremental addition of multiple MQTT broker nodes.

Based on experimental evidence (Table 4, Table 5 and Table 6), we built a platform to address some of the key technical challenges organisations face when building a new IoT application, including:
(1) reliable and scalable business-critical IoT application;
(2) high data throughput to meet expectations of the end-user for responsible IoT products;
(3) lower cost of operation through efficient hardware network and cloud resources;
(4) integrating IoT data into existing TBP platform.

# 4 DISCUSSION

In the case of the testbed platform for Czech Post, the proposed solution had to meet various requirements intended to modernise the existing eSystem. In general, IoT data streaming and horizontally scalable analytical platforms intended to advance eCommerce and eGovernment systems can be evaluated in terms of achieved performance, security, interoperability, and data increase before the subsequent horizontal scaling-up eSystem reconfiguration.

Regarding the estimate on how many virtual clients should be connected to the MQTT broker to generate up to 150.000 messages, we found that ten clients/represented as IoT devices worked well and without any performance degradation.

As a limitation of this study, we have:
- Experimented on a single node MQTT broker.
- Set the maximum number of generated and processed messages (up to 150.000 messages/second).
- Set the zero-tolerance for message loss (not requested by the client) since we assumed there might be no 'Acknowledgement' response between some sensor network devices and the edge gateway.
- Not reported the technical details of Middleware component, developed as the proprietary solution, which is non-disclosable due to commercial arrangements and expectations between the authors and the Czech Post as our client.
- Within the scope of the project and horizontally scalable architecture, we did not emulate large number of IoT devices.

Regarding the single node MQTT broker configuration that could be categorised as a low-cost 'commodity hardware' (see Table 1), we reported:
(1) How many sensor devices was handled in emulations without horizontally scaling up the platform; and
(2) The number of messages per second, without experiencing message loss during our experiments.

As a result of multiple "cold start" experiments (required to avoid experimental bias associated with memory caching), we have determined that 150.000 is the number of messages per second where there is no message loss. Contrary to our expectations, there was no non-linear decline in message throughput performance (Figure 4).

We have also considered the possibility of messaging broker components by testing HiveMQ VerneMQ and Mosquitto but have only presented the benchmark for HiveMQ on a single virtual node as an edge gateway. The VerneMQ and Mosquitto brokers were excluded from the reported case study due to lower performance. Note that Mosquitto broker is not designed for distributed multi-node MQTT broker operation.

The added contribution of this study is developed and shared as an open-source Java application (https://github.com/stufim/mqtt-performance). For this and related projects, the Java application is designed to be connected to the testbed platform architecture running on a single node MQTT broker for platform benchmarking, using an assigned IP address with the default port of 1883. This application can also be used for multi-node distributed MQTT brokers for similar projects without any major changes. The results demonstrate that HiveMQ showed the best throughput with no message lost in our scenario. Moreover, it allows analysing the interplay between docker container and container orchestration on Kubernetes (K8s).

We found that: (1) IoT sensors connected to the messaging broker via MQTT protocol affect its performance. For example, MQTT explorer (http://mqtt-explorer.com) slows down performance by about 30% (depending on which Apache Kafka's topic has been connected to as a client). (2) TBP can handle and proceed up to 150.000 messages per second for a limited time (less than a minute) on HiveMQ messaging broker in real-time.

The paper presents an overall design of a streaming framework aligned with the predefined parameters for benchmarking real-time data processing. To ensure (near) real-time data exchange and processing on the testbed platform, with data loss, the individual components (Figure 2) are responsible for: (1) event streaming, transformation, and processing, (2) data storage to collect, store and manage analytical processing requests and (3) application layer. In addition, the middleware component handles business logic, object models, and session control (sockets and APIs) as the application layer.

The presentation layer was intended for a web browser-based client written in Java framework (which in our case is not to be disclosed due to commercial sensitivity). However, aligned with reported literature, the presented concept, experimental design, and insights presented in this case study are transferrable to other eSystems design and development.

## 5 CONCLUSION AND FUTURE WORK

For this study, we designed a testbed platform to meet the Czech Post's requirements for the horizontally scalable analytical platform and messaging throughput rates from various IoT devices by using multi-thread Java application. The presented performance evaluation results confirm that the requirement of two million messages per 24 hours has been met. Experimental evidence confirmed that the presented system can handle peaks of 150.000 IoT messages per second (SMS and emails) on a single node MQTT broker and without any message loss.

The case study reported the minimum viable platform for real or near real-time streams data processing using HiveMQ broker by using multi-thread Java application (Figure 5). VerneMQ and Mosquitto MQTT brokers were excluded due to their lower performance measured in the initial experimental test scenarios. Apart from the developed middleware component for our client (the Czech Post), all other components were based on or provided as open-source software.

We have also demonstrated the appropriateness and transferability of experimental design applicable to similar real-time processing analytical platforms. Furthermore, the presented platform applies to a range of IoT sensor network contexts.

As an added value, this study also presented an open-source multi-thread Java application for benchmarking a real-time testbed distributed platform for IoT streaming data (Solutia, 2021). Regarding how to use a multi-thread Java application, available as an open-source software download from GitHub (https://github.com/stufim/mqtt-performance), the XML file (mqtt-performance/pom.xml) is provided to facilitate configuration of Java libraries, dependencies, and version control. Furthermore, this source code applies to similar and extended development contexts involving sensor networks to single-node and distributed MQTT streaming platforms.

Future work will provide a solution to support the large number of clients (in order of millions). Depending on the Czech Post preference, future work is also likely to consider:

(1) integration with well-established analytical software used for business intelligence (such as Tableau, Qlik and Microsoft's Power BI);

(2) integration with identity management and security enhancement tools (such as CAS Protocol for Single Sign-On and other Network monitoring and Application Performance tools); and

(3) distributed processing scenarios with multiple-node performance measures and other technologies such as EMQX broker, RabbitMQ and ActiveMQ messaging systems that could be used for integration via APIs from other components of the Czech Post eSystem.

## ACKNOWLEDGEMENTS

The data collection, benchmarks and the testbed platform setup required considerable collaborative efforts from Solutia, s.r.o. (Prague) employees. We are grateful for their contribution and effort during the benchmarking and setting up infrastructure for these purposes.

# APENDIX

Figure 5: GitHub libraries (https://github.com/stufim/mqtt-performance) import for multi-thread Java application PublishThread.java.

# VOCABULARY

| | |
|---|---|
| Commodity Hardware | Relatively inexpensive, off-the-shelf, non-proprietary, interchangeable, consumer-grade computers or electronic devices intended for building parallel computing infrastructures. |
| Internet of Things (IoT) | Network-attached devices that can typically exchange data processed from connected or embedded sensors. In this study, we focus on IoT relying on Wi-Fi and Internet Protocol. |
| Sensor | A device that can quantify inputs from detected changes, or distinguish properties or events of occurring phenomena of interest, and share acquired input with other information-processing infrastructures. |
| Sensor Network | A group of small devices connected in one or more networks. |
| Wireless Device | A device that doesn't require physical connection for communication. |